\documentclass[conference,letterpaper]{IEEEtran}

\usepackage[utf8]{inputenc} 
\usepackage[T1]{fontenc}
\usepackage{url}
\usepackage{ifthen}
\usepackage{cite}
\usepackage{amssymb}
\usepackage[cmex10]{amsmath} 
\usepackage{algorithm} 
\usepackage{algorithmic}

\newtheorem{theorem}{Theorem}[section]
\newtheorem{proposition}[theorem]{Proposition}

\begin{document}
\title{Weak Superimposed Codes of Improved Asymptotic Rate and Their Randomized Construction}

\author{%
  \IEEEauthorblockN{Yu Tsunoda}
  \IEEEauthorblockA{Faculty of Engineering, Information and Systems\\
                    University of Tsukuba\\
                    Tsukuba 305-8573, Japan\\
                    Email:  y.tsunoda@sk.tsukuba.ac.jp}
  \and
  \IEEEauthorblockN{Yuichiro Fujiwara}
  \IEEEauthorblockA{Division of Mathematics and Informatics\\
                    Chiba University\\ 
                    1-33 Yayoi-Cho Inage-Ku, Chiba 263-8522, Japan\\
                    Email: yuichiro.fujiwara@chiba-u.jp}
}

\maketitle

\begin{abstract}
Weak superimposed codes are combinatorial structures related closely to generalized cover-free families, superimposed codes, and disjunct matrices in that they are only required to satisfy similar but less stringent conditions.
This class of codes may also be seen as a stricter variant of what are known as locally thin families in combinatorics.
Originally, weak superimposed codes were introduced in the context of multimedia content protection against illegal distribution of copies under the assumption that
a coalition of malicious users may employ the averaging attack with adversarial noise.
As in many other kinds of codes in information theory, it is of interest and importance in the study of weak superimposed codes to find the highest achievable rate in the asymptotic regime and give an efficient construction that produces an infinite sequence of codes that achieve it.
Here, we prove a tighter lower bound than the sharpest known one on the rate of optimal weak superimposed codes and give a polynomial-time randomized construction algorithm for codes that asymptotically attain our improved bound with high probability.
Our probabilistic approach is versatile and applicable to many other related codes and arrays.
\end{abstract}


\section{Introduction}\label{intro}
In information theory, a binary code of length $l$ is usually a finite set of $l$-dimensional vectors over the binary field $\mathbb{F}_2$ with specific properties.
For instance, binary linear codes of length $l$, which form one of the most fundamental classes of codes in coding theory, are simply subspaces of the $l$-dimensional vector space $\mathbb{F}_2^l$.

The special properties a code is required to possess naturally depend on its intended application.
From this viewpoint, digital communications and many of the other fields where information theory is prominently applied to are typically discrete in nature.
Hence, those properties which are considered in coding theory tend to be combinatorial.
Most typically, desirable codes correspond to extremal sets that have as many elements as possible while meeting a certain set of combinatorial requirements. With the \textit{rate} of a binary code $\mathcal{C}$ of length $l$ defined as the fraction of the binary logarithm $\log_2(\vert\mathcal{C}\vert)$ of the size of $\mathcal{C}$ relative to the dimension $l$ of $\mathbb{F}_2^l$, we are often interested in the highest possible rate of a code that satisfies particular properties.

As is evident from the fact that coding theory and combinatorics have flourished together in information theory, intriguing applications of codes often require those that satisfy interesting properties from the viewpoint of combinatorics.
This fortunate tendency can, however, lead to situations where essentially the same mathematical objects are being studied independently in coding theory, combinatorics, and those fields that require interesting codes without realizing each others' research.

\textit{Superimposed codes} and their variants are examples of such repeated rediscoveries among classic codes.
This type of code was originally introduced for data retrieval systems in 1964 \cite{Kautz1964} and has subsequently found other applications in coding theory from multiple-access communications \cite{Bonis2003, Wolf1985, Berger1984} to digital fingerprinting such as those found in \cite{Egorova2020, Barg2003, Boneh1998},
leading to extensive theoretical investigations into superimposed codes and their generalizations and close variants. For very recent examples, see, for example, \cite{Fan2021, Gargano2020, Gu2018} and references therein.
Superimposed codes are also known as \textit{cover-free families} in combinatorics, where the same combinatorial objects were initially studied in \cite{Erdos1985, Erdos1982} from a purely mathematical viewpoint and followed by other combinatorially oriented research (see, for example, \cite{Stinson2004, Yeh2002, Ruszinko1994} and references therein).
Yet another name for a superimposed code is a \textit{disjunct matrix}, which frequently appears in the context of non-adaptive group testing \cite{Du2006, Du2000} and has long been an important concept since its invention in \cite{Bush1984}.

While independent lines of research on the equivalent notion in different fields can be a source of major breakthroughs in respective disciplines, studies on the same topic from different angles may place their focuses slightly differently.
For instance, coding theorists and combinatorial theorists may both be interested in the asymptotic behavior of the rate of largest codes given other parameters.
However, coding theory regards efficient construction algorithms as equally or even more important when the codes under consideration have practical applications, while not as much emphasis has been placed on this aspect for cover-free families and related structures in combinatorics.
In this regard, the constructive aspect is undoubtedly vital in non-adaptive group testing, regardless of which variant of a disjunct matrix is considered.

The purpose of this paper is to develop a versatile method that allows for both proving the existence of and providing an efficient construction algorithm for a wide variety of codes with large rates that are all similar to superimposed codes.
To best illustrate our general idea in a succinct manner, we specifically consider \textit{weak superimposed codes}, which were recently introduced in the context of multimedia content protection \cite{Egorova2020}.
Weak superimposed codes are particularly suited for illuminating the advantages of our simple idea partly because
their weak structural requirement has prevented obtaining a strong rate bound through well-known other approaches
and also partly because their intended application strongly requires a constructive, not just existential, proof for the existence of desired codes.
Indeed, the original motivation behind weak superimposed codes is to provide an efficient algorithm for finding malicious parties who illegally distribute multimedia content.

In this paper, through a simple and general probabilistic approach, we improve the best known lower bound on the largest achievable rate of weak superimposed codes given in \cite{Egorova2020} and provide a polynomial-time randomized algorithm for constructing codes that asymptotically attain our improved bound with high probability.

As will be defined formally in the next section, in addition to the length and size, a weak superimposed code has two more parameters $t$ and $d$, which are assumed to be positive constants.
As coding-theoretic objects, it is of natural interest to analyze the asymptotic code rate $R(t,d)$ of a largest code given other parameters.
While results from combinatorics can reveal that $R(t,d)$ is upper bounded by roughly $\frac{2}{t}$, we will prove the following.
\begin{theorem}\label{deletion}
For any integers $t\ge 2$ and $d \ge 1$,
\begin{align*}
R(t,d)\ge  \frac{\log_2 e}{t-1}\left(1-\frac{1}{t}\right)^{t-1},
\end{align*}
where $e$ is the base of the natural logarithm.
\end{theorem}
This is an improvement on the tightest known bound by a factor of $\frac{t}{t-1}$ and also a constructively achievable bound through a randomized construction.
While the parameter $d$ is usually assumed to be constant, mathematically speaking, our proof remains valid if $d$ depends on $l$ as long as it grows slowly enough as $l$ tends to infinity.

While the above result may just be another instance of an elementary probabilistic method leading to an efficient construction,
for related codes to which no equally strong techniques have successfully been applied,
our approach may improve the sharpest known lower bounds on the highest achievable rates and give a new construction method.
This is the case with special set systems known as \textit{generalized cover-free families} \cite{Stinson2004}, where the tightest corresponding general bounds are given in \cite{Bui2021, Deng2004}.
For similar codes to which sophisticated tools such as the algorithmic Lov\'{a}sz Local Lemma \cite{Moser2010} have already been applied effectively,
our method may give an alternative proof for the strongest known bounds and another efficient construction algorithm.
A related class of codes, called X-\textit{codes} \cite{Tsunoda2018, Fujiwara2010, Mitra2004, Lumetta2003, Kong2021, Tsunoda2019, Tsunoda2018a}, supplies an example of this case, where our method leads to a bound which is asymptotically the same as the tightest known one given in \cite{Tsunoda2018, Tsunoda2017} and provides us with an alternative efficient construction.

It should be noted, however, that our approach does not always result in the best known bound.
The most important caveat in this regard applies to the very original problem on the rate of superimposed codes of the vanilla kind,
where our method can only prove a somewhat weaker bound than the tightest known one obtained through a very complicated probabilistic analysis of binary constant-weight codes \cite{Dyachkov1989}.
Therefore, our contribution should be viewed as a general approach which is sufficiently powerful to prove a strong bound on the highest achievable rate and sufficiently simple to allow for extracting an efficient randomized construction algorithm for a wide range of codes related to superimposed codes.

In the following section, we formally define weak superimposed codes and give a brief review of relevant results.
Section \ref{mainr} gives a probabilistic proof of an improved lower bound on the largest achievable rate of weak superimposed codes
and explains how to extract a randomized construction algorithm from our proof.
Section \ref{conclude} discusses how our method may be applied to other similar codes and concludes this paper with some remarks.

\section{Preliminaries}\label{pre}
Here, we formally define a weak superimposed code and give a brief review of known results.
We first present a coding-theoretic definition by following the original paper \cite{Egorova2020} and then give two more equivalent ones for the reader who is interested in related results found in the literature in the language of set systems and/or matrices.
Throughout this paper, $e$ is the base of the natural logarithm.

A \textit{weak $(t,d)$-superimposed code} of \textit{length} $l$ and \textit{size} $n$ is a subset $\mathcal{C} \subseteq \mathbb{F}_2^l$ of the $l$-dimensional vector space over $\mathbb{F}_2$ such that for any subset $S \subseteq \mathcal{C}$ with $2\le |S|\le t$, there are at least $d$ coordinates $i$ such that
the number $\pi_i(S)$ of elements $(c_0, \dots, c_{l-1}) \in S$ in which the $i$th coordinate is $1$ is exactly $1$, that is, for any $t$-subsets $S \subseteq \mathcal{C}$, at least $d$ coordinates $i$ satisfy $\pi_i(S) = \vert\{(c_0, \dots, c_{l-1}) \in S\mid c_i=1\}\vert = 1$.
The parameter $t$ is the \textit{strength} of the weak superimposed code $\mathcal{C}$,
while each vector in $\mathcal{C}$ is a \textit{codeword}.
As usual, the \textit{support} of a vector $\boldsymbol{v} = (v_0, \dots, v_{l-1}) \in \mathbb{F}_2^l$ is the set $\operatorname{supp}(\boldsymbol{v}) = \{i \mid v_i = 1\}$ of nonzero coordinates of $\boldsymbol{v}$.
The cardinality of the support of $\boldsymbol{v}$ is the \textit{weight} of the vector.

As is the case with many other codes related to superimposed codes, weak superimposed codes can naturally be defined in the language of set systems as well.
A \textit{set system} of \textit{order} $l$ is an ordered pair $(V, \mathcal{B})$ of a set $V$ of cardinality $l$ and a family $\mathcal{B}$ of subsets of $V$.
Without loss of generality, we assume that $V = \{0, 1, \dots, l-1\}$.
The elements of $V$ are the \textit{points}, while those of $\mathcal{B}$ are \textit{blocks}.
Given a set system $(V, \mathcal{B})$ of order $l$ with exactly $n$ blocks,
by regarding the $l$ points as the coordinates of the $l$-dimensional vector space $\mathbb{F}_2^l$ and
the $n$ blocks as the supports of corresponding $n$ vectors in $\mathbb{F}_2^l$,
it is straightforward to see that $(V, \mathcal{B})$ is a weak $(t,d)$-superimposed code of length $l$ and size $n$ if and only if
for any subset $\mathcal{S} \subseteq \mathcal{B}$ with $2 \leq \vert \mathcal{S}\vert \leq t$, there exist at least $d$ points each of which is contained in exactly one block in $\mathcal{S}$.

To view a weak superimposed code as a special matrix, consider the point-by-block incidence matrix $H = (h_{i,j})$ of the corresponding set system $(V, \mathcal{B})$, where the rows and columns are indexed by the points and blocks, respectively, and $h_{i,j} = 1$ if the point $i \in V$ is in the block $j \in \mathcal{B}$ and $0$ otherwise.
From this point of view, it is easy to see that an $l \times n$ matrix over $\mathbb{F}_2$ forms a weak $(t,d)$-superimposed code of length $l$ and size $n$ if and only if for any $l \times s$ submatrix with $2 \leq s \leq t$, there exist at least $d$ rows of weight $1$.

As is pointed out in \cite{Egorova2020}, for $t = 2$, a weak $(2,d)$-superimposed code is just a binary code of minimum Hamming distance at least $d$,
which can be verified straightforwardly from the coding-theoretic definition.
For $t \geq 3$, however, the two notions differ because weak superimposed codes must satisfy not just a pairwise requirement but also $s$-wise one among codewords for any $2 \leq s \leq t$.
For other notions whose special cases coincide with weak superimposed codes of particular kind, we refer the reader to \cite{Egorova:2021ti}.

From a coding-theoretic viewpoint, we are mainly concerned with the asymptotic behavior of the highest achievable \textit{rate} $\frac{\log_2(\vert\mathcal{C}\vert)}{l}$ of a weak $(t,d)$-superimposed code $\mathcal{C}$ for fixed $t$ and $d$ as the length $l$ tends to infinity.
Let $F(t, d; l)$ be the maximum size of a weak $(t,d)$-superimposed code of length $l$ and define the \textit{asymptotic code rate} $R(t, d)$ of extremal $(t,d)$-weak superimposed codes by
\begin{equation*}
R(t, d)=\underset{l\rightarrow\infty}{\overline{\lim}}\frac{\log_2 F(t, d;l)}{l}.
\end{equation*}

The sharpest known bound on $R(t, d)$ for general $t$ and $d$ is given through a version of the basic probabilistic argument with the union bound.
\begin{theorem}[\cite{Egorova2020}]\label{original}
For any integers $t\ge 2$ and $d\ge 1$,
\begin{equation*}
R(t, d)\ge \frac{\log_2 e}{t}\left(1-\frac{1}{t}\right)^{t-1}.
\end{equation*}
\end{theorem}
Note that the right-hand side of the inequality in Theorem \ref{original} can be bounded from below by $\frac{\log_2e}{et}$ without losing its tightness too much.
It is also notable that, while the above theorem was proved in \cite{Egorova2020} in a way a randomized algorithm produces a weak superimposed code asymptotically attaining the above bound with probability at least $\frac{1}{2}$, it is straightforward to modify the proof so that the success probability becomes any positive constant less than $1$.

A nontrivial general upper bound on $R(t, d)$ can be obtained by borrowing a result from combinatorics.
For positive integer $t \geq 2$, a set system $(V, \mathcal{B})$ is $t$-\textit{locally thin} if for any subset $\mathcal{S} \subseteq \mathcal{B}$ of exactly $t$ blocks, at least one point $v \in V$ is contained in exactly one block in $\mathcal{S}$.
Clearly, any weak $(t,d)$-superimposed code is a $t$-locally thin set system because the former is just a set system that is simultaneously $s$-locally thin for all $2 \leq s \leq t$.

While there does not seem to be any mention of the following upper bound on $R(t, d)$ in the literature, this can be obtained straightforwardly by invoking the known result on $t$-locally thin set systems.
\begin{theorem}\label{upper}
For any integers $t\ge 2$ and $d\ge 1$,
\begin{align*}
R(t, d) < \frac{1}{\left\lfloor\frac{t}{2}\right\rfloor}.
\end{align*}
\end{theorem}
\begin{IEEEproof}
Let $G(u; l)$ be the maximum size of a $u$-locally thin set system of order $l$.
By Corollary 3.1 in \cite{Alon2000}, for any even $u \geq 2$, we have
\begin{align*}
\underset{l\rightarrow\infty}{\overline{\lim}}\frac{\log_2 G(u; l)}{l} < \frac{2}{u}.
\end{align*}
Because a weak $(t,d)$-superimposed code is $s$-locally thin for all $2 \leq s \leq t$, the assertion follows.
\end{IEEEproof}
It is notable that if we do not require our upper bound to be in closed form, the above theorem can be strengthened by invoking Theorem 2.1 in \cite{Alon2000} instead.
However, the resulting bound would be considerably more complicated with a limited improvement.
For instance, through numerical computation over a complicated inequality, we can show that $R(4, d) < 0.4968$ for any $d \geq 1$ with this approach.
The interested reader is referred to \cite{Bonis2019, Alon2000} and references therein.

\section{Main Results}\label{mainr}
We first prove our main existence theorem.
Our proof strategy is a simple application of the alteration method in probabilistic combinatorics \cite{Alon2016},
which is also known as the sample-and-modify method \cite{Mitzenmacher2017}.
For convenience, we restate our main theorem here.
\setcounter{section}{1}
\setcounter{theorem}{0}
\begin{theorem}
For any integers $t\ge 2$ and $d \ge 1$,
\begin{align*}
R(t,d)\ge  \frac{\log_2 e}{t-1}\left(1-\frac{1}{t}\right)^{t-1}.
\end{align*}
\end{theorem}
\setcounter{section}{3}
\setcounter{theorem}{0}
\begin{IEEEproof}
We view a weak superimposed code as a special matrix.
Let $t$, $d$, $l$, and, $n$ be positive integers such that $n > t \geq 2$ and $l > d \geq 1$.
Take an $l\times n$ random matrix $C=(c_{i,j})$ over $\mathbb{F}_2$ by defining each entry $c_{i,j}$ to be $1$ with probability $p$ and $0$ with probability $1-p$ independently and uniformly at random.
Let $\mathcal{A}$ be the set of subsets $A$ of columns in $C$ with $2 \le \vert A \vert \le t$.
For $A\in\mathcal{A}$, let $C_A$ be the $l \times \vert A\vert $ submatrix of $C$ which consists of the columns in $A$
and define $E_A$ to be the event that $C_A$ has at most $d-1$ rows of weight $1$.

Let $X_i$ be the indicator random variable such that
\begin{equation*}
X_A=\begin{cases}
1 & \text{if event }E_A \text{ occurs,}\\
0& \text{otherwise}
\end{cases}
\end{equation*}
and define $X=\sum_{A\in\mathcal{A}}X_A$, which counts the number of subsets $A$ of columns in $\mathcal{C}$ such that 
the corresponding $l \times \vert A\vert$ submatrix $C_A$ has at most $d-1$ rows of weight $1$.
Deleting at most $X$ columns chosen appropriately from $C$ gives a $(t, d)$-superimposed code of length $l$ and size at least $n-X$.
Therefore, if
\begin{align*}
\mathbb{E}(X)<\frac{n}{2},
\end{align*}
then there exists a $(t, d)$-superimposed code of length $l$ and size at least $n'=\frac{n}{2}$.

We bound $\mathbb{E}(X)$ from above.
Define $\mathcal{A}_s$ to be the set of subsets $A$ of columns in $C$ with $\vert A\vert = s$
and let $q(s,p) = sp(1-p)^{s-1}$.
For any $A \in\mathcal{A}_s$, the probability $\Pr(E_A)$ that the event $E_A$ occurs is 
\begin{align*}
\Pr(E_A) &= \sum_{i=0}^{d-1}\binom{l}{i}q(s,p)^i(1-q(s,p))^{l-i}\\
&=\left(1-q(s,p)\right)^l\sum_{i=0}^{d-1}\binom{l}{i}\left(\frac{q(s,p)}{1-q(s,p)}\right)^i.
\end{align*}
Set $p=\frac{1}{t}$. Because $0< q\!\left(s,\frac{1}{t}\right) \le \frac{1}{2}$ for any $2\le s\le t$, we have
\begin{equation*}
\frac{q\!\left(s,\frac{1}{t}\right)}{1-q\!\left(s,\frac{1}{t}\right)} \le 1.
\end{equation*}
Hence, by linearity of expectation, we have
\begin{align*}
\mathbb{E}(X)&=\sum_{A\in\mathcal{A}}\mathbb{E}(X_A)\\
&= \sum_{s=2}^t\sum_{A\in\mathcal{A}_s}\mathbb{E}(X_A)\\
&= \sum_{s=2}^t\sum_{A\in\mathcal{A}_s}\Pr(X_A)\\
&<l^d\sum_{s=2}^tn^s\!\left(1-q\!\left(s,\frac{1}{t}\right)\right)^l\sum_{i=0}^{d-1}\left(\frac{q\!\left(s,\frac{1}{t}\right)}{1-q\!\left(s,\frac{1}{t}\right)}\right)^i\\
&\leq dl^d\sum_{s=2}^t n^s\left(1-q\!\left(s,\frac{1}{t}\right)\right)^l.
\end{align*}

Therefore, if it holds that
\begin{align}\label{bad_p1}
 n^s\left(1-q\!\left(s,\frac{1}{t}\right)\right)^l \le \frac{n}{2(t-1)dl^d}
\end{align}
for any $2\le s\le t$, then there exists a weak $(t,d)$-superimposed code of length $l$ and size at least $n'$.
Taking the logarithm of both sides of Inequality (\ref{bad_p1}) and diving by $l$ gives
\begin{align*}
\frac{s-1}{l}\log_2 n +\log_2\!\left(1-q\!\left(s,\frac{1}{t}\right)\right)&\le -\frac{\log_2\!\left(2(t-1)dl^d\right)}{l}.
\end{align*}
Hence, because $\ln (1-x)<-x$ for $0<x<1$ and $n'=\frac{n}{2}$, if 
\begin{align*}
\frac{\log_2 n'}{l} \le \frac{s\log_2 e}{(s-1)t}\!\left(1-\frac{1}{t}\right)^{s-1}-\frac{\log_2\!\left( 2(t-1)dl^d\right)}{(s-1)l}-\frac{1}{l},
\end{align*}
then Inequality (\ref{bad_p1}) holds.
Now, the right-hand side of the above inequality satisfies the following inequality.
\begin{align*}
 \frac{s\log_2 e}{(s-1)t}&\left(1-\frac{1}{t}\right)^{s-1}-\frac{\log_2\!\left( 2(t-1)dl^d\right)}{(s-1)l}-\frac{1}{l}\nonumber\\
& \ge  \frac{t\log_2 e}{(t-1)t}\left(1-\frac{1}{t}\right)^{t-1}-\frac{\log_2 \!\left(2(t-1)dl^d\right)}{l}-\frac{1}{l}.\label{altrate}
\end{align*}
Therefore, the following inequality
\begin{align*}
\frac{\log_2 n'}{l}& \le \frac{\log_2 e}{t-1}\!\left(1-\frac{1}{t}\!\right)^{t-1}-\frac{\log_2 \left(2(t-1)dl^d\right)}{l}-\frac{1}{l}
\end{align*}
is a sufficient condition for the existence of a weak $(t, d)$-superimposed code of length $l$ and size at least $n'$.
Hence, we have
\begin{equation*}
 R(t, T)\ge \frac{\log_2 e}{t-1}\left(1-\frac{1}{t}\right)^{t-1}
\end{equation*}
as desired.
\end{IEEEproof}

Next, we extract an efficient randomized construction algorithm from the above proof of Theorem \ref{deletion}.
The idea is that because a random binary code of length $l$ in an appropriate probability space considered in the proof contains exponentially many codewords, throwing away a fraction of codewords does not affect its asymptotic rate.
Hence, while we started with an $l \times n$ random matrix and deleted at most $\mathbb{E}(X) = \frac{n}{2}$ columns in our probabilistic proof,
the coefficient $\frac{1}{2}$ in the expectation can be replaced with an arbitrary constant without sacrificing its rate in the asymptotic sense.
This implies that a simple application of Markov's inequality suffices for driving the probability of producing a desired weak superimposed code arbitrarily close to $1$.
\begin{proposition}[Markov's inequality]
Let $X$ be a nonnegative random variable and $a>0$.
Then, it holds that
\begin{equation*}
\Pr(X\ge a)\le \frac{\mathbb{E}(X)}{a}.
\end{equation*}
\end{proposition}
Therefore, as in the probabilistic proof of Theorem \ref{deletion}, all we have to do is start with an $l \times n$ random matrix $H=(h_{i,j})$ over $\mathbb{F}_2$ of appropriate size with each entry $h_{i,j} = 1$ independently and uniformly at random with probability $p = \frac{1}{t}$, look for $l \times s$ submatrices with $2 \leq s \leq t$ in which fewer than $d$ rows are of weight $1$, and delete them all by discarding a column from each such submatrix.

For positive $f = o(2^{l})$, let
\begin{align*}
r(t,d,l,f) = \frac{\log_2 e}{t-1}\!\left(1-\frac{1}{t}\!\right)^{t-1}\!-\frac{\log_2 \left(2f(t-1)dl^d\right)}{l}-\frac{2}{l}.
\end{align*}
A set $C \subseteq \mathbb{F}_2^l$ of codewords of size $2\le |C|\le t$ is $(t,d)$-\textit{violated} if there exists at most $d-1$ coordinates $i$ such that $\pi_i(C)=1$.
Our randomized construction algorithm is described in Algorithm \ref{delalgo}, where a weak superimposed code is viewed as a special binary code.

\begin{algorithm}[H]
    \caption{Construction for a weak superimposed code}
    \label{delalgo}
    \begin{algorithmic}[1]
    \REQUIRE Integers $t\ge 2$, $d\ge 1$, and $l>0$, and real $f > 0$
    \ENSURE weak $(t, d)$-superimposed code $\mathcal{C}$
    \STATE $\mathcal{C}\leftarrow$ $\lceil2^{r(t, d, l, f)l+1}\rceil$ random vectors $\boldsymbol{c} \in \mathbb{F}_2^l$, where each coordinate of $\boldsymbol{c}$ is $1$ independently with probability $\frac{1}{t}$
    \WHILE{$\exists$ $(t,d)$-violated $C\subseteq\mathcal{C}$ with $2\le |C|\le t$}
    \STATE Arbitrarily pick $(t,d)$-violated $C$ and its codeword $\boldsymbol{c}$
    \STATE $\mathcal{C}\leftarrow \mathcal{C}\setminus \{\boldsymbol{c}\}$
    \ENDWHILE
    \RETURN $\mathcal{C}$
    \end{algorithmic}
\end{algorithm}

\begin{theorem}\label{deletion_prob}
Let $t \geq 2$ and $d \geq 1$.
For any positive $f = o(2^{l})$,
Algorithm \ref{delalgo} outputs a weak $(t, d)$-superimposed code of rate at least $r(t, d, l, f)$ with probability at least $1-f^{-1}$,
where
\begin{align*}
r(t,d,l,f) &= \frac{\log_2 e}{t-1}\!\left(1-\frac{1}{t}\!\right)^{t-1}\!-\frac{\log_2 \left(2f(t-1)dl^d\right)}{l}-\frac{2}{l}\\
&\rightarrow \frac{\log e}{t-1}\left(1-\frac{1}{t}\right)^{t-1}
\end{align*}
as $l \rightarrow \infty$, and runs in time polynomial in the size of the code as a binary matrix.
\end{theorem}
\begin{IEEEproof}
Let $n=\lceil2^{r(t,d,l,f)l+1}\rceil$.
As in the proof of Theorem \ref{deletion}, take an $l \times n$ random matrix $C=(c_{i,j})$ over $\mathbb{F}_2$ by defining each entry $c_{i,j}$ to be $1$ with probability $\frac{1}{t}$ and $0$ with probability $1-\frac{1}{t}$ independently and uniformly at random.
By Markov's inequality, the probability $\Pr(X \geq f\mathbb{E}(X))$ that the number $X$ of $(t,d)$-violated sets of codewords is larger than or equal to its expectation $\mathbb{E}(X)$ satisfies
\begin{align*}
\Pr(X \geq f\mathbb{E}(X)) \leq f^{-1}.
\end{align*}
Note that if $t$, $d$, $l$, $n$, $f$ are such that $\mathbb{E}(X) < \frac{n}{2f}$, we have
\begin{align*}
\Pr\left(X < \frac{n}{2}\right) \geq 1-f^{-1}.
\end{align*}
Thus, following the same argument as in the proof of Theorem \ref{deletion},
it is routine to check that, with probability at least $1-f^{-1}$, Algorithm \ref{delalgo} outputs a weak $(t,d)$-superimposed code of length $l$ and size at least $n'$, where
\begin{align*}
\frac{\log_2 n'}{l} &= \frac{\log_2 e}{t-1}\!\left(1-\frac{1}{t}\!\right)^{t-1}-\frac{\log_2 \left(2f(t-1)dl^d\right)}{l}-\frac{2}{l}\\
&\rightarrow \frac{\log_2 e}{t-1}\!\left(1-\frac{1}{t}\!\right)^{t-1}
\end{align*}
as $l$ tends to infinity.

It now suffices to show that the algorithm runs in time polynomial in the number of codewords.
Picking a random code $\mathcal{C}$ takes $O(ln)$ time with $l=O(\log n)$, while at most $\sum_{s=2}^t\binom{n}{s}=O(n^{t})$ subcodes are required to be checked in the while loop.
Since each check takes at most $O(\log n)$ time, the total running time is $O(n^t\log n)$.
\end{IEEEproof}

\section{Concluding remarks}\label{conclude}
We have derived a lower bound on the highest achievable rate of weak superimposed codes in the asymptotic regime.
Our proof of Theorem \ref{deletion} is a natural application of the alteration method in probabilistic combinatorics,
where the simplicity of our approach allowed for converting the probabilistic argument into a polynomial-time construction algorithm.

Our method is applicable to some other similar codes whose characteristics are coordinate-wise properties among small sets of codewords.
For instance, for positive $w$, $r$, and $d$, a set system $(V, \mathcal{B})$ is called a $(w,r;d)$-\textit{cover-free family} if for any disjoint pair $\mathcal{X}, \mathcal{Y} \subseteq \mathcal{B}$ of sets of blocks with $\vert \mathcal{X}\vert = w$ and $\vert \mathcal{Y}\vert = r$, it holds that
\begin{align*}
\left\vert \bigcap_{A \in \mathcal{X}}A \ \middle\backslash \ \bigcup_{B \in \mathcal{Y}}B\right\vert \geq d.
\end{align*}
A $(1,r;1)$-cover-free family is equivalent to an $r$-superimposed code and hence forms a weak $(r,1)$-superimposed code as well.
In terms of asymptotic rate, the general and strongest known lower bounds roughly say that it is at least $\frac{w^wr^r}{8(w+r-1)(w+r)^{w+r}}$ in \cite{Deng2004}
and at least $\frac{w^wr^r}{(w+r)^{w+r+1}}$ in \cite{Bui2021}.
Both results invoke the Chernoff bound, while the former also relies on the Lov\'{a}sz Locall Lemma.
Here, it is routine to verify that the straightforward application of the alteration method as we did in the proof of Theorem \ref{deletion} is powerful enough to show that it is asymptotically bounded from below by $\frac{w^wr^r}{(w+r-1)(w+r)^{w+r}}$, which improves the best known bounds.
Because Algorithm \ref{delalgo} straightforwardly follows the alteration method, essentially the same randomized construction works also for $(w,r;d)$-cover-free families except that the definition of violation of a partial structure which appears in the while loop needs a slight modification.

It is worth noting that the algorithmic Lov\'{a}sz Local Lemma in symmetric form and the naive alteration approach seem to often provide asymptotically the same bounds for combinatorial arrays similar to those we considered in this paper.
This was noted in \cite{Deng2004}, where various combinatorial arrays were studied in a unified fashion.
As yet another example, this phenomenon can be observed also by applying our approach to X-codes which we briefly mentioned earlier in Section \ref{intro}.
Roughly speaking, an $(l,n,t,x)$ X-code is a set of $n$ codewords taken from $\mathbb{F}_2^l$ such that for every positive integer $s \leq t$, no superimposed sum of any $x$ codewords covers the vector obtained by adding up any $s$ codewords chosen from the rest of the $n-x$ codewords.
The highest achievable rate is proven to be asymptotically at least $\frac{\log_2 e}{e(x+1)(x+t-1)}$ by the symmetric Lov\'{a}sz Local Lemma \cite{Tsunoda2018} as well as the alteration method \cite{Tsunoda2017}.
It is not difficult to derive an efficient randomized constriction algorithm asymptotically meeting this bound through the approach we took in this paper.

Interestingly, weak superimposed codes appear somewhat different from some other codes we have seen in this paper in that the naive application of the symmetric Lov\'{a}sz Local Lemma does not seem to prove as tight a bound in this case.
While we have not ruled out the possibility that a more technical application of the Lov\'{a}sz Local Lemma is as effective for weak superimposed codes,
it seems also plausible that codes which are only required to meet certain types of weak conditions pose their own unique challenges to theoretical investigations.
It is hoped that further research overcomes those obstacles and makes significant progress on notoriously difficult problems around codes, set systems, and combinatorial arrays.

\section*{Acknowledgment}
The authors thank the anonymous referees for their valuable comments.
This work was supported by JSPS KAKENHI Grant Number JP21J00593 (Y.T.) and KAKENHI Grant Number JP20K11668 (Y.F.).





\end{document}